\documentclass[12pt]{article}
\usepackage{amssymb,amsmath,amsthm,amsfonts,amscd}
 \usepackage{graphicx}
\newcommand\be{\begin{equation}}
\newcommand\ee{\end{equation}}
\newcommand\bea{\begin{eqnarray}}
\newcommand\eea{\end{eqnarray}}

\newcommand{\fatalpha}{{\bf \alpha \kern -0.44em \alpha}}
\newcommand{\fatsigma}{{\bf \sigma \kern -0.54em \sigma}}
\newcommand{\tpchi}{{\bf D \kern -0.35em D}}
\newcommand{\llambda}{{\bf \lambda \kern -0.45em \lambda}}

\renewcommand{\theequation}{\arabic{equation}}
\renewcommand{\theequation}{\thesection.\arabic{equation}}
\bibliography{plain}
\title{\bf Quantum discord of  $2^{n}$-dimensional Bell-diagonal states }\vspace{20mm}
\author{ M. A. Jafarizadeh $^{a}$
 \thanks{E-mail:jafarizadeh@tabrizu.ac.ir}  ,
  N. Karimi  $^{b}$
 \thanks{E-mail:na\_karimi@yahoo.com} ,
 D. Amidi $^{a}$
 \thanks{E-mail:daavudamidi@gmail.com}
  and
  H. Zahir  $^{a}$
 \thanks{E-mail:Zahir.sci@gmail.com}
\\ $^a${\small Department of Theoretical Physics and Astrophysics,
Tabriz University, Tabriz 51664, Iran.} \\ $^b${\small Farhangian University,  PardisAllameh Amini ,Tabriz, Iran.}} \pagebreak

\vspace{20mm}
\begin{document}
\maketitle \vspace{15mm}
\newpage
\begin{abstract}
 In this study, using the concept of relative entropy as a distance measure of correlations we investigate the important issue of evaluating quantum correlations such as entanglement, dissonance and classical correlations for $2^{n}$-dimensional Bell-diagonal states. We provide an analytical technique, which describes how we find the closest classical states(CCS) and the closest separable states(CSS) for these states. Then analytical results are obtained for quantum discord of $2^{n}$-dimensional Bell-diagonal states. As illustration, some special cases are examined. Finally, we investigate the additivity relation between the different correlations for the separable generalized bloch sphere states.
\end{abstract}

{\bf PACs Index: 03.67.-a, 03.65.Ta, 03.65.Ud}
\vspace{70mm}
\newpage
\section{Introduction}
Quantum entanglement plays an important role in the
quantum communication protocols like teleportation \cite{Bennett,Horodecki},
superdense coding \cite{Wiesner}, remote state preparation \cite{Pati}, cryptography
\cite{Gisin} and many more.
However, entanglement is not the only correlation that is useful for
quantum information processing. Recently, it is found
that many tasks, e.g. quantum nonlocality without entanglement \cite{R,C,J}, can be carried out with quantum
correlations other than entanglement. It has been shown
both theoretically and experimentally \cite{A,B} that some
separable states may speed up certain tasks over their
classical counterparts. Recent measures of nonclassical correlations are motivated by different notions of classicality and operational
means to quantify nonclassicality. One kind of nonlocal correlation called quantum discord, as introduced by Oliver and Zurek \cite{Z1,ved}, has received
much attention recently \cite{Z2,Z3,Z4,Z5,Z6,Z7,Z8,Z9,Z10,Z11,Z12,Z13,Z14}. Most of these
works are limited to studies of bipartite correlations only as the concept of discord, which relies on the definition of
mutual information, is not defined for multipartite systems. It is well known that the different measures of quantum correlation
are not identical and conceptually different. For example, the discord does not coincide with entanglement and
a direct comparison of two notions is rather meaningless. Therefore, a unified classification of correlations is in demand.
 Modi et. al.\cite{Z8}, introduced a unified classification of correlations for quantum states which is applicable for multipartite systems. In this unified view, the  measure of correlation is based on the idea that the distance from a given state to the closest state without the desired property (e.g. entanglement or discord) is a measure of that property. Finding the CCS is still a very
difficult problem and has the same challenged as faced in computing original discord. \cite{rana1,rana2}.
The examples of entangled states $\rho$ with analytical expression for the CSSs, discussed in Refs.\cite{a1, a2, a8, a9, a10, a11, a15, a16,a17}.
The inverse problem to the long standing problem \cite{11} of finding the formula for the CSS was solved in \cite{12} for the case of two qubits and
a closed formula for all entangled states was solved in \cite{13} for all dimensions and for any number of parties. In this paper, we give an efficient procedure so that analytic evaluation of quantum discord of $2^{n}$-dimensional Bell-diagonal states can be performed. Then we find an exact explicit
formula for quantum discord of these states. We also show that total correlation for the separable generalized bloch sphere states is subadditive.\\
This paper is organized as follows. In the next section we introduce the distance measure of correlation and show that the generic state $\rho$ and its CSS or CCS have the same structure. In section 3 the definition of $2^{n}$-dimensional Bell-diagonal states is given and in
section 4 we calculate the CCS of $2^{n}$-dimensional Bell-diagonal states and then we find an exact analytical
formula for the quantum discord of these states. In the rest of this section to illustrate how the formula can be applied, we give two examples.
In section 5 we investigate additivity relations between different correlations for separable generalized bloch sphere states. Concluding remarks and two Appendices close this paper.

\section{Distance measure of correlations.}
Here, we will follow the approach of \cite{Z8} to characterize and quantify all kinds of correlations in a quantum state. The definitions of relevant quantities are:
\begin{equation}\label{eehg}
    Entanglement \quad E=\min_{\sigma\in\mathcal{ D}} S(\rho\|\sigma),
\end{equation}
\begin{equation}\label{d1}
    Discord \quad D=\min_{\chi\in \mathcal{C}} S(\rho\|\chi),
\end{equation}
\begin{equation}\label{q1}
    Dissonance \quad Q=\min_{\chi\in \mathcal{C}} S(\sigma\|\chi),
\end{equation}
\begin{equation}\label{q51}
   Classical \ correlations \quad C=\min_{\pi\in \mathcal{P}} S(\chi\|\pi),
\end{equation}
where $\mathcal{P}$ is the set of all product states (i.e., states of the
form $\pi=\pi_{1}\otimes\pi_{2}\otimes...\otimes\pi_{N}$ and $\pi_{n}$ is the reduced state of the nth subsystem). $\mathcal{C}$ contains mixtures of locally distinguishable states $\chi=\sum_{k_{n}}p_{k_{1}...k_{N}}|k_{1}...k_{N}\rangle\langle k_{1}...k_{N}|=\sum_{\vec{k}}p_{\vec{k}}
|\vec{k}\rangle\langle\vec{k}|$ where $p_{\vec{k}}$ is a joint probability distribution and local states $|k_{n}\rangle$ span an orthonormal basis, $\mathcal{ D}$ is the set of all separable states (i.e., states of the form $\sigma=\sum_{k}p_{k}\pi_{1}^{k}\otimes\pi_{2}^{k}\otimes...\otimes\pi_{N}^{k}$)
and $S(x\|y)=Tr(x\log x-x\log y)$ is the relative entropy of x with respect to y.\\
A bipartite state is called classical if it contains mixtures of locally distinguishable states, and is called separable if it can be represented as a convex combination of product states. Finding out the CSS is a non trivial task\cite{rana2}. While the set of separable states is apparently convex, this is not the case for the set of classical states then determining the CCS is even more complicated.
Here we present an analytical procedure that allows us to obtain the CSS and CCS for $2^{n}$-dimensional Bell-diagonal states.
The key idea is to find  the minimum distance from a given state $\rho$ to the set of all  states without the desire property. The following theorem plays a central role in minimizing the mentioned distance.\\
\textbf{Theorem:}
 Given a generic state $\rho \in \mathcal{H}\backslash \mathcal{I}$ and $X \in \mathcal{I}$, min $S(\rho\|X)$ is achieved when  $\rho$ and $X$ have common eigenbasis. Here, $\mathcal{I}$ is a special subset of the Hilbert space $\mathcal{H}$.\\
 To show this, suppose
 $$
 \rho=\sum_{i}^{N} \lambda_{i}|\lambda_{i}\rangle\langle \lambda_{i}|,\quad X=\sum_{j}^{N} \mu_{j}|\mu_{j}\rangle\langle \mu_{j}|,
 $$
 then we have
 \begin{equation}\label{lag8}
 \min S(\rho\|X)=\min[Tr\rho \log \rho- Tr(\rho \log X)]=\sum\lambda_{i}\log\lambda_{i}-\max \sum_{i,j}\lambda_{i}|\langle \lambda_{i}|\mu_{j}\rangle|^{2}\log \mu_{j}
 \end{equation}
 Suppose $|\langle \lambda_{i}|\mu_{j}\rangle|^{2}=q_{ij}$ , where
 \begin{equation}\label{lkkp2}
\quad \sum_{i}q_{ij}=1, \quad \sum_{j}q_{ij}=1.
 \end{equation}
 Now the problem of finding the closest state $X$ to $\rho$ is reduced to the problem
 \begin{equation}\label{lp22}
\left\{ \begin{array}{c}
\rm{maximize}\quad \sum_{i,j}\lambda_{i}q_{ij}\log \mu_{j}=\lambda^{T}Q\eta .\\
\rm{subject\; to}\quad
 \quad \sum_{i}q_{ij}=1, \quad \sum_{j}q_{ij}=1\\
\end{array}\right.
\end{equation}
 where $\eta^{T}=(\log \mu_{1},\log \mu_{2},...,\log \mu_{N^{2}})$, $\lambda^{T}=(\lambda_{1},\lambda_{2},...,\lambda_{N^{2}})$\\

 Eq. (\ref{lkkp2}) shows that the matrix $Q$ with the $(Q)_{ij}=q_{ij}$ is doubly stochastic matrix.
 The set of doubly stochastic matrices, $\Omega_{n}$, is the convex hull of the
permutation matrices (Birkhoff (1946), von Neumann (1953)). In other words,
the doubly stochastic matrix, $\Omega_{n}$, is the convex combination
of the permutation matrices, $P_{n}$, that is
\begin{equation}\label{lpp2}
Q= \sum_{i}\tau_{i}P_{i}\quad ,\sum_{i}\tau_{i}=1,\quad \tau_{i}\geq 0 \quad \forall i
\end{equation}
so, Eq.(\ref{lp22}) takes form
\begin{equation}\label{lkp2}
\rm{maximize}\quad \sum_{i}\tau_{i}\lambda^{T}P_{i}\eta  ,\quad \sum_{i}\tau_{i}=1,\\
\end{equation}
hence our problem reduces to a Linear Programming optimization  over the convex set of feasible region. Here the feasible region is a simplex and
  its apex, yield when one of the $\tau_{i}$ equals 1 and the others equal zero, are the desired solutions of this optimization problem. This means that
$\rho$ and $X $ have common eigenbasis.\\
\section{Definition of $2^{n}$-dimensional Bell-diagonal states}
In order to put our discussion in a precise setting, let us first introduce $2^{n}$-dimensional Bell-diagonal states acting on a bipartite system $H^{A}\otimes H^{B}$ with $dim(H^{A})=N = 2^{n}$ and $dim(H^{B})=N = 2^{n}$.
 To do this, let $S=<g_{1}=\gamma_{1} \otimes\gamma_{1},...,g_{2n}=\gamma_{2n} \otimes\gamma_{2n}\}>$ be generated by 2n independent and commuting element such that  -I is not an element of S and  $g_{i}^{2}=I$ for all $g_{i}\in S$ .\\
 $\gamma_{j}$ for j =1, 2, . . . , 2n+1, known as Dirac matrices. (For a brief review about Dirac matrices
and an explicit construction of $\gamma_{j}$, see Appendix II).\\
Hence, we can represent the density operators acting on a bipartite system $H^{A}\times H^{B
 }$ as:
  \begin{equation}\label{jj}
  \rho=\frac{1}{N^{2}}\sum_{i_{1},i_{2},...,i_{2n}}t_{i_{1},i_{2},...,i_{2n}}g^{i_{1}}_{1}g^{i_{2}}_{2}...g^{i_{2n}}_{2n}
 \end{equation}
 where $i_{1},i_{2},...,i_{2n} \in \{0,1\}$\\
 Consider the projection operators $\{\pi_{i_{1},i_{2},...,i_{2n}}\}$ with
 $$
 \pi_{i_{1},i_{2},...,i_{2n}}=\frac{1}{2^{2n}}\prod_{j=1}^{2n}(I+(-1)^{i_{j}}g_{j})=
 \sum_{j_{1},j_{2},...,j_{2n}}^{1}(-1)^{i_{1}j_{1}+i_{2}j_{2}+...+i_{2n}j_{2n}}g_{1}^{j_{1}}g_{2}^{j_{2}}...g_{2n}^{j_{2n}}
 $$
 \begin{equation}\label{p}
\pi_{i_{1},i_{2},...,i_{2n}}\pi_{j_{1},j_{2},...,j_{2n}}=\delta_{i_{1}j_{1}}...\delta_{i_{2n}j_{2n}}\pi_{i_{1},i_{2},...,i_{2n}}, \quad \sum_{i_{1},i_{2},...,i_{2n}}  \pi_{i_{1},i_{2},...,i_{2n}}=I,\\
\end{equation}
then we get
 \begin{align}\label{ro}
  \rho=\sum_{i_{1},i_{2},...,i_{2n}} p_{i_{1},i_{2},...,i_{2n}}\pi_{i_{1},i_{2},...,i_{2n}}
 \end{align}
 where
$$
 p_{i_{1},i_{2},...,i_{2n}}=\frac{1}{N^{2}}\sum_{j_{1},j_{2},...,j_{2n}}(-1)^{i_{1}j_{1}+i_{2}j_{2}+...+i_{2n}j_{2n}}t_{j_{1},j_{2},...,j_{2n}}
 $$
 \begin{equation}\label{phj}
t_{j_{1},j_{2},...,j_{2n}}=\sum_{i_{1},i_{2},...,i_{2n}}(-1)^{i_{1}j_{1}+i_{2}j_{2}+...+i_{2n}j_{2n}}p_{i_{1},i_{2},...,i_{2n}}\\
\end{equation}
From the theorem above it follows that the CSS states can be represented as:
\begin{equation}\label{sig}
  \sigma=\frac{1}{N^{2}}\sum_{i_{1},i_{2},...,i_{2n}}\acute{t}_{i_{1},i_{2},...,i_{2n}}g^{i_{1}}_{1}g^{i_{2}}_{2}...g^{i_{2n}}_{2n}=\sum_{i_{1},i_{2},...,i_{2n}} \acute{p}_{i_{1},i_{2},...,i_{2n}}\pi_{i_{1},i_{2},...,i_{2n}}
  \end{equation}
  and CCS states are
  $$
  \chi_{\rho}=\frac{1}{N^{2}}\sum_{i_{1},i_{2},...,i_{2n}}\tilde{t}_{i_{1},i_{2},...,i_{2n}}g^{i_{1}}_{1}g^{i_{2}}_{2}...g^{i_{2n}}_{2n}=\sum_{i_{1},i_{2},...,i_{2n}} q_{i_{1},i_{2},...,i_{2n}}\pi_{i_{1},i_{2},...,i_{2n}}
 $$
 \begin{equation}\label{sig}
  \chi_{\sigma}=\frac{1}{N^{2}}\sum_{i_{1},i_{2},...,i_{2n}}\bar{t}_{i_{1},i_{2},...,i_{2n}}g^{i_{1}}_{1}g^{i_{2}}_{2}...g^{i_{2n}}_{2n}=\sum_{i_{1},i_{2},...,i_{2n}} \acute{q}_{i_{1},i_{2},...,i_{2n}}\pi_{i_{1},i_{2},...,i_{2n}}
  \end{equation}
\section{Calculation of classical states and quantum discord}
First of all, we note that the result of above theorem can straightforwardly be used to obtain the CCS states of  $2^{n}$-dimensional Bell-diagonal states. To do this recall that the Pauli operators on a single qubit are $\{I,\sigma_{x},\sigma_{y},\sigma_{z}\}$. The representation of the Pauli group we will deal with is the group formed by elements of the form $G_{n}=\{i^{k}P_{1}\otimes P_{2}\otimes...\otimes P_{n}\}$ where each $P_{i}$ is an element of   $\{I,\sigma_{x},\sigma_{y},\sigma_{z}\}$.

 Suppose $\Gamma$ is a subgroup of $G_{n}$ generated by elements $\{\Gamma_{i_{1}i_{2}...i_{n}j_{1}j_{2}...j_{n}}\}$. There is an extremely useful way of presenting the generators of $\Gamma$ \cite{boyd}. To do this, we use $r(\Gamma_{i_{1}i_{2}...i_{n}j_{1}j_{2}...j_{n}})=[i_{1}i_{2}...i_{n}|j_{1}j_{2}...j_{n}]$ to denote the 2n-dimensional row vector representation of an element of the $\Gamma$. The left hand side of the row vector contains 1s to indicate which generators contain $\sigma_{x}$s, and the right hand side contains 1s to indicate which generators contain  $\sigma_{z}s$; the presence of a 1 on both sides indicates a  $\sigma_{y}$ in the generator. More explicitly, it is constructed as follows. If $\gamma_{i}$ contains an I on the jth qubit then the jth and n+jth column elements are 0; if it contains an  $\sigma_{x}$ on the jth qubit then the jth column element is a 1 and the n+jth column element is a 0; if it
contains a  $\sigma_{z}$ on the jth qubit then the jth column element is 0 and the n+jth column element is 1; if it contains a  $\sigma_{y}$ on the jth qubit
then both the jth and n+jth columns are 1. Let us define a $2n\times2n$ matrix $\Lambda$ by
 \begin{equation}\label{phfdj}
\left(
  \begin{array}{cc}
    0 & I_{n\times n}\\
    I_{n\times n} & 0 \\
  \end{array}
\right),
\end{equation}
then the elements $\Gamma_{i_{1}i_{2}...i_{n}j_{1}j_{2}...j_{n}}$ and $\Gamma_{\acute{i}_{1}\acute{i}_{2}...\acute{i}_{n}\acute{j}_{1}\acute{j}_{2}...\acute{j}_{n}}$ are easily seen to commute if and only if
\begin{equation}\label{hfdlj}
r(\Gamma_{i_{1}i_{2}...i_{n}j_{1}j_{2}...j_{n}})\Lambda r(\Gamma_{\acute{i}_{1}\acute{i}_{2}...\acute{i}_{n}\acute{j}_{1}\acute{j}_{2}...\acute{j}_{n}})= 0 \ ( mod\ 2 ).
\end{equation}
Let $g_{c_{k}}\in\{\Gamma_{i_{1}i_{2}...i_{n}j_{1}j_{2}...j_{n}}\otimes\Gamma_{i_{1}i_{2}...i_{n}j_{1}j_{2}...j_{n}}\}$ for $k=1,...,n$ , such that the Eq.(\ref{hfdlj}) is satisfied, then we can rewrite the state $\rho$ such as:
 $$
  \rho=\frac{1}{N^{2}}\sum_{i_{1},...,i_{2n}}t_{i_{1},...,i_{2n}}g_{c_{1}}^{i_{1}}...g_{c_{n}}^{i_{n}}g_{n+1}^{i_{n+1}}...g_{2n}^{i_{2n}}.
  $$
 Since classical states contains mixtures of locally distinguishable
states hence we can rewrite the expression of the CCS of $\rho$ such as:
  \begin{equation}\label{133}
 \chi=\frac{1}{N^{2}}\sum_{i_{1},...,i_{2n}}\tilde{t}_{i_{1},i_{2},...,i_{n},\underbrace{0,0,...,0}_{nfold}}
 g_{c_{1}}^{i_{1}}...g_{c_{n}}^{i_{n}}g_{n+1}^{0}...g_{2n}^{0}=\frac{1}{N^{2}}\sum_{i_{1},...,i_{2n}}
 \tilde{t}_{i_{1},i_{2},...,i_{n},\underbrace{0,0,...,0}_{nfold}}
 g_{c_{1}}^{i_{1}}...g_{c_{n}}^{i_{n}}
 \end{equation}
 where $t_{\underbrace{0,...,0}_{nfold}}=\tilde{t}_{\underbrace{0,...,0}_{nfold}}=1$.\\
 Here we show that some of $\tilde{t}_{i_{1},i_{2},...,i_{n},\underbrace{0,0,...,0}_{nfold}}$ are zero and rest of them are the same of $t_{i_{1},i_{2},...,i_{n},\underbrace{0,0,...,0}_{nfold}}$.
To show this note that the eigenvalues of $\chi$ which are  $2^{n}-$fold degenerate, are given by
 \begin{equation}\label{1133}
  q_{i_{1},...,i_{n},i_{n+1},...,i_{2n}}=\frac{1}{N^{2}}\sum_{j_{1}
  ,...,j_{n}}(-1)^{i_{1}j_{1}+...+i_{n}j_{n}}\tilde{t}_{j_{1},j_{2},...,j_{n},\underbrace{0,0,...,0}_{nfold}},
  \end{equation}
 so the problem of finding the CCS to $\rho$ is reduced to the problem
 \begin{equation}\label{lp2}
\left\{ \begin{array}{c}
\rm{minimize}\quad S(\rho\|\chi)\\
\rm{subject\; to}\quad \sum_{i_{1},...,i_{n},i_{n+1},...,i_{2n}}q_{i_{1},...,i_{n},i_{n+1},...,i_{2n}}=1.\\
\end{array}\right.
\end{equation}
 The dual Lagrangian associated with this problem, is given by
$$
 L=\sum_{i_{1},...,i_{2n}} p_{i_{1},...,i_{2n}}\log p_{i_{1},...,i_{2n}}-\sum_{i_{1},...,i_{n}}(\sum_{i_{n+1},...,i_{2n}}p_{i_{1},...,i_{n},i_{n+1},...,i_{2n}})\log q_{i_{1},...,i_{n},i_{n+1},...,i_{2n}}
$$
\begin{equation}\label{11343}
+\mu[\sum_{i_{1},...,i_{n},i_{n+1},...,i_{2n}}q_{i_{1},...,i_{n},i_{n+1},...,i_{2n}}-1].
\end{equation}
By calculating the gradient of the dual Lagrangian with respect to $\tilde{t}_{j_{1},j_{2},...,j_{n},\underbrace{0,0,...,0}_{nfold}}$ and making it zero we get
\begin{equation}\label{3438}
\frac{\partial L}{\partial \tilde{t}_{j_{1},j_{2},...,j_{n},\underbrace{0,0,...,0}_{nfold}}}=-\sum_{i_{1},...,i_{n}}[\frac{\sum_{i_{n+1},...,i_{2n}}p_{i_{1},...,i_{n},i_{n+1},...,
i_{2n}}}{q_{i_{1},...,i_{n},i_{n+1},...,i_{2n}}}+\mu](-1)^{i_{1}j_{1}+...+i_{n}j_{n}}=0.
\end{equation}
Since
\begin{equation}\label{88}
\sum_{i_{n+1},...,i_{2n}}p_{i_{1},...,i_{n},i_{n+1},...,i_{2n}}=\frac{2^{n}}{N^{2}}\sum_{j_{1},j_{2},...,j_{2n}}(-1)^{i_{1}j_{1}+...+i_{n}j_{n}}t_{j_{1},j_{2},...,j_{n},\underbrace{0,0,...,0}_{nfold}},
\end{equation}
then using Eq. (\ref{3438})  one can show that
\begin{equation}\label{797}
\tilde{t}_{j_{1},j_{2},...,j_{n},\underbrace{0,0,...,0}_{nfold}}=t_{j_{1},j_{2},...,j_{n},\underbrace{0,0,...,0}_{nfold}} \quad \forall j_{1},...,j_{n},
\end{equation}
then we can rewrite the CCS of $\rho$ such as:
  \begin{equation}\label{13e3}
 \chi=\frac{1}{N^{2}}\sum_{i_{1},...,i_{2n}}
 t_{i_{1},i_{2},...,i_{n},\underbrace{0,0,...,0}_{nfold}}
 g_{c_{1}}^{i_{1}}...g_{c_{n}}^{i_{n}}
 \end{equation}
In the rest of this section we calculate the quantum discord of $2^{2n}$-dimensional Bell-diagonal states. Note that the key difference between the original definition of discord\cite{Z1,ved} and the definition in Eq.(\ref{d1}) is in minimization. We minimize the quantity D, while for the original discord, $D-L_{\rho}$ is minimized \cite{Z8} where $L_{\rho}=S(\pi_{\chi_{\rho}})-S(\pi_{\rho})$ . Since for the Bell-diagonal states $L_{\rho}=0$, hence the two forms of discord are the same. Now, using Eq.(\ref{797}) we give an exact analytical formula quantum discord for $2^{n}$-dimensional Bell-diagonal states such as:
$$
D=\sum_{i_{1},i_{2},...,i_{2n}} p_{i_{1},i_{2},...,i_{2n}}\log p_{i_{1},i_{2},...,i_{2n}}-\max\frac{1}{N^{2}}\sum_{i_{1},i_{2},...,i_{n}}\sum_{j_{1}
  ,...,j_{n}}(-1)^{i_{1}j_{1}+...+i_{n}j_{n}}t_{j_{1},j_{2},...,j_{n},\underbrace{0,0,...,0}_{nfold}}
  $$
 \begin{equation}\label{discord}
\log \frac{1}{N^{2}}\sum_{j_{1},...,j_{n}}(-1)^{i_{1}j_{1}+...+i_{n}j_{n}}t_{j_{1},j_{2},...,j_{n},\underbrace{0,0,...,0}_{nfold}}
\end{equation}
where the maximum is taken over all parameters $\{t_{j_{1},j_{2},...,j_{n},\underbrace{0,0,...,0}_{nfold}}\}$.
      Here we should mention that the set $\{g_{c_{1}}^{i_{1}}...g_{c_{n}}^{i_{n}}\}$ in Eq.(\ref{13e3}) can be chosen in many different ways. Since in the optimum strategy D is at its minimum, then we chose the set $\{g_{c_{1}}^{i_{1}}...g_{c_{n}}^{i_{n}}\}$ or equivalently the parameters $\{t_{j_{1},j_{2},...,j_{n},\underbrace{0,0,...,0}_{nfold}}\}$ such that the second part in the Eq.(\ref{discord}) is maximized. To give an intuitive understanding of this subject, let us illustrate it by a fundamental examples.\\
\subsection{Example 1: Generalized bloch sphere states}
 Using Eq.(\ref{jj}) the generalized bloch sphere states \cite{khodam} are given by
 \begin{equation}\label{v}
 \rho=\frac{1}{N^{2}}[I+\sum_{k=1}^{2n}t_{0,...,\underbrace{1}_{k},...,0}g_{k}+ t_{1,1,...,1}g_{2n+1}]
 \end{equation}
 where $g_{2n+1}=g_{1}g_{2}...g_{2n}$.
Since classical states contains mixtures of locally distinguishable states hence using Eq.(\ref{13e3})
one can conclude that the CCS of generalized bloch sphere states lie on the Cartesian axes. That is only one of the $\{t_{i_{1},i_{2},...,i_{n},\underbrace{0,0,...,0}_{nfold}}\}$ is nonzero.
    We assume that, without loss of generality, $t_{1,0,...,0}\neq 0$, then the CCS of generalized bloch sphere states are given by
   \begin{equation}\label{l}
   \chi=\frac{1}{N^{2}}[I+t_{1,0,...,0}g_{1}],
   \end{equation}
In the optimum strategy $S(\rho\|\chi)$ is at its minimum, that is we have $t_{1,0,...,0}= t_{max}$. Hence using, (\ref{discord}), we obtain

 \begin{align}\label{nha2}
D=\sum_{i_{1},i_{2},...,i_{2n}}^{N^{2}-1}p_{i_{1},i_{2},...,i_{2n}} \log p_{i_{1},i_{2},...,i_{2n}}-\frac{1-t_{max}}{2}\log(1-t_{max})-
\frac{1+t_{max}}{2}\log(1+t_{max})+2\log(N).
 \end{align}
 This is in agreement with the result obtained in \cite{khodam} for N=M.
\subsection{Example 2: $2^{2}$-dimensional Bell-diagonal states}
 As the second example, to keep our discussion simple, let us focus attention on n=2 case.
In this case, using (\ref{13e3}) we have
\begin{equation}\label{1323}
 \chi=\frac{1}{16}\sum_{i_{1},...,i_{4}}t_{i_{1},i_{2},0,0}
 g_{c_{1}}^{i_{1}}g_{c_{2}}^{i_{2}}
 \end{equation}
 where, using (\ref{hfdlj}) we get
\begin{equation}\label{hgbfdj}
\begin{array}{ccc}
( g_{c_{1}},g_{c_{2}})\in\{(\Gamma_{1000}\otimes\Gamma_{1000},\Gamma_{0101}\otimes \Gamma_{0101}),(\Gamma_{0001}\otimes \Gamma_{0001},\Gamma_{0110}\otimes \Gamma_{0110}),(\Gamma_{1000}\otimes \Gamma_{1000},\Gamma_{0110}\otimes \Gamma_{0110}),\\
      (\Gamma_{0100}\otimes \Gamma_{0100},\Gamma_{0011}\otimes \Gamma_{0011}),(\Gamma_{1111}\otimes \Gamma_{1111},\Gamma_{1100}\otimes \Gamma_{1100}),( \Gamma_{1111}\otimes \Gamma_{1111},\Gamma_{1010}\otimes \Gamma_{1010}),\\

   (\Gamma_{1000}\otimes \Gamma_{1000},\Gamma_{0011}\otimes \Gamma_{0011}),(\Gamma_{0010}\otimes \Gamma_{0010},\Gamma_{0101}\otimes \Gamma_{0101}),(\Gamma_{1111}\otimes \Gamma_{1111},\Gamma_{1001}\otimes \Gamma_{1001}), \\

    (\Gamma_{0010}\otimes \Gamma_{0010},\Gamma_{1001}\otimes \Gamma_{1011}),(\Gamma_{0100}\otimes \Gamma_{0100},\Gamma_{1010}\otimes \Gamma_{1010}),(\Gamma_{0001}\otimes \Gamma_{0001},\Gamma_{1100}\otimes \Gamma_{1100})  \\

   (\Gamma_{0010}\otimes \Gamma_{0010},\Gamma_{1100}\otimes \Gamma_{1100}),(\Gamma_{0100}\otimes \Gamma_{0100},\Gamma_{1001}\otimes \Gamma_{1001}), (\Gamma_{0001}\otimes \Gamma_{0001},\Gamma_{1010}\otimes \Gamma_{1010})\}, \\
      \end{array}
\end{equation}
and
\begin{equation}\label{hgbdj}
\begin{array}{ccc}
  \Gamma_{1000}=\sigma_{x}\otimes I,\Gamma_{0100}=i\sigma_{y}\otimes \sigma_{x},\Gamma_{0010}=i\sigma_{y}\otimes \sigma_{y},
   \Gamma_{0001}=i\sigma_{y}\otimes \sigma_{z},\Gamma_{1111}=-\sigma_{z}\otimes I,\\
   \Gamma_{1100}=-\sigma_{z}\otimes \sigma_{x},\Gamma_{1010}=-\sigma_{z}\otimes \sigma_{y},\Gamma_{1001}=i\sigma_{z}\otimes \sigma_{z},\Gamma_{0110})=-iI\otimes \sigma_{z} ,\Gamma_{0101}=I\otimes \sigma_{y},\\
   \Gamma_{0011}=-I\otimes \sigma_{x},\Gamma_{0111}=\sigma_{x}\otimes I, \Gamma_{1011}=\sigma_{x}\otimes \sigma_{x},\Gamma_{1101}=\sigma_{x}\otimes \sigma_{y},\Gamma_{1110}=\sigma_{x}\otimes \sigma_{z}.
\end{array}
\end{equation}
Let us choose $g_{c_{1}}=\Gamma_{1000}\otimes\Gamma_{1000}$ and $g_{c_{2}}=\Gamma_{0101}\otimes \Gamma_{0101}$, hence we have
\begin{equation}\label{kapa}
\chi=\frac{1}{16}(I+t_{1000}\Gamma_{1000}\otimes \Gamma_{1000}+t_{0100}\Gamma_{0101}\otimes \Gamma_{0101}+t_{1100}\Gamma_{1101}\otimes \Gamma_{1101})
\end{equation}
  Then, using (\ref{discord}) quantum discord is given by
\begin{equation}\label{dis11}
 D=\sum_{i_{1},i_{2},i_{3},i_{4}}p_{i_{1},i_{2},i_{3},i_{4}}\log p_{i_{1},i_{2},i_{3},i_{4}}-\sum_{i=1}^{4}p_{q_{i}}\log p_{q_{i}}.
 \end{equation}
   where
 \begin{equation}\label{44}
 \begin{array}{c}
   p_{q_{1}}=\frac{1}{16}[1+t_{1000}+t_{0100}+t_{1100}]\\
   p_{q_{2}}=\frac{1}{16}[1+t_{1000}-t_{0100}-t_{1100}]  \\
   p_{q_{3}}=\frac{1}{16}[1-t_{1000}+t_{0100}-t_{1100}] \\
   p_{q_{4}}=\frac{1}{16}[1-t_{1000}-t_{0100}+t_{1100}]
 \end{array}
 \end{equation}
  \section{Subadditivity of correlations of a quantum state}
  It has been conjectured \cite{Z8} that the correlations of a quantum state are subadditive in the sense $T_{\rho}\geq E+Q+C_{\sigma}$ (where $T_{\rho}$ is total mutual information which defined as $S(\rho\|\pi_{\rho})$ and $C_{\sigma}$ is the classical
correlation $S(\chi_{\sigma}|\pi_{\sigma})$).
  In general, from an analytical point of view, the derivation of closed expressions of relative entropy of entanglement involves optimization procedures that are very complicated to perform. Hence, here we consider the inverse problem \cite{12} and investigate additivity relations between different correlations for separable generalized bloch sphere states. These states are given by
   \begin{equation}\label{rosep2}
 \sigma=\frac{1}{N^{2}}[I+\sum_{k=1}^{2n}\acute{t}_{0,...,\underbrace{1}_{k},...,0}g_{k}+
 \acute{t}_{1,1,...,1}g_{2n+1}],
 \end{equation}
 with the eigenvalues
 \begin{align}\label{rsep22}
 \acute{p}_{i_{1},i_{2},...,i_{2n}}=\frac{1}{N^{2}}[1+\sum_{k=1}^{2n}(-1)^{i_{k}}
 \acute{t}_{0,...,\underbrace{1}_{k},...,0}+(-1)^{n}(-1)^{i_{1}+...+i_{2n}}\acute{t}_{1,1,...,1}]
 \end{align}
  where $g_{2n+1}=g_{1}g_{2}...g_{2n}$.
The separable generalized bloch sphere states are actually bounded by
$\sum_{k=1}^{2n}|\acute{t}_{0,...,\underbrace{1}_{k},...,0}|+|\acute{t}_{1,1,...,1}|\leq 1 $
or, equivalently, $\acute{p}_{i_{1},i_{2},...,i_{2n}} \leq \frac{2}{N^{2}}$ $(\forall i_{1},i_{2},...,i_{2n})$.
   The family of all entangled states, $\rho(x,\sigma)$ , for which $\sigma$ is the CSS is given by \cite{13}
 \begin{align}\label{rosep22}
 \rho(x,\sigma)=\sigma-xL_{\sigma}^{-1}(w_{i_{1},i_{2},...,i_{2n}}), \quad 0<x\leq x_{max}
 \end{align}
 where $L_{\alpha}$ is linear operator. In the eigenbasis of $\alpha$, $\alpha=diag(a_{1},...,a_{n})$ is a diagonal
matrix, where $a_{1},...,a_{n}> 0$ and for any $\beta=[b_{i,j=1}^{n}]$, $L_{\alpha}(\beta)$
is defined by
\begin{align}\label{beta}
[ L_{\alpha}(\beta)]_{kl}=\left\{ \begin{array}{c}
b_{kl}\frac{\ln a_{k}-\ln a_{l}}{a_{k}-a_{l}}, \quad if \quad a_{k}\neq a_{l} \\
 b_{kl}\frac{1}{a}, \quad if \quad a_{k}= a_{l}=a .\\
\end{array}\right.
 \end{align}
  $ w_{i_{1},i_{2},...,i_{2n}}$,s are entanglement witnesses (EW) of $2^{n}$-dimensional Bell-diagonal states.
 Here, $x_{max}$ is defined such that $\rho(x_{max},\sigma)\in H_{n,+,1}$(convex set of positive hermitian
matrices of trace one) and
$\rho(x_{max},\sigma)$ has at least one zero eigenvalue. We also note that $w_{i_{1},i_{2},...,i_{2n}}$ is normalized; i.e. $Tr(w_{i_{1},i_{2},...,i_{2n}})^{2}=1$ and $Tr(L_{\sigma}^{-1}(w_{i_{1},i_{2},...,i_{2n}}))=0$. In general EW  of the $2^{2n}$-dimensional Bell-diagonal states is given by \cite{jafarizadeh1}
\begin{align}\label{ew}
 w_{i_{1},i_{2},...,i_{2n}}=\frac{1}{N\sqrt{2(n+1)}}[I_{2^{2n}}+\sum_{k=1}^{2n}(-1)^{i_{k}}\gamma_{k}^{2n} \otimes \gamma_{k}^{2n}-(-i)^{2n}(-1)^{i_{1}+i_{2}+...+i_{2n}}\gamma_{2n+1}^{2n} \otimes \gamma_{2n+1}^{2n}],
 \end{align}
 with the eigenvalue
\begin{align}\label{ewegen}
 \lambda_{j_{1},j_{2},...,j_{2n}}^{w_{i_{1},i_{2},...,i_{2n}}}=
 \frac{1}{N\sqrt{2(n+1)}}[1+\sum_{k=1}^{2n}(-1)^{i_{k}}(-1)^{j_{k}}
 -(-1)^{i_{1}+i_{2}+...+i_{2n}+j_{1}+j_{2}+...+j_{2n}}].
 \end{align}
  Now define the real symmetric matrix \cite{13}
 \begin{align}\label{s}
[S(\sigma)]_{k_{1}...k_{2n},l_{1}...l_{2n}}=\frac{\acute{p}_{k_{1}...k_{2n}}-\acute{p}_{l_{1}...l_{2n}}}{\ln \acute{p}_{k_{1}...k_{2n}}-\ln \acute{p}_{l_{1}...l_{2n}}},
 \end{align}
 hence we get
 \begin{align}\label{ss}
S(\sigma)=\sum_{k_{1}...k_{2n}\neq l_{1}...l_{2n}}\frac{\acute{p}_{k_{1}...k_{2n}}-\acute{p}_{l_{1}...l_{2n}}}{\ln \acute{p}_{k_{1}...k_{2n}}-\ln \acute{p}_{l_{1}...l_{2n}}}\Pi_{k_{1}...k_{2n}}\Pi_{l_{1}...l_{2n}}+\sum_{k_{1}...k_{2n}}\acute{p}_{k_{1}...k_{2n}}\Pi_{k_{1}...k_{2n}}.
 \end{align}
 Note that $L_{\sigma}$ is an invertible operator, where $L_{\sigma}^{-1}(w_{i_{1},i_{2},...,i_{2n}})=w_{i_{1},i_{2},...,i_{2n}}\bullet S(\sigma)$
where $A\bullet B$ is the entrywise product of two matrices of A and B.
 Let
  \begin{align}\label{sss}
w_{i_{1},i_{2},...,i_{2n}}=\sum_{j_{1}...j_{2n}}\lambda_{j_{1}...j_{2n}}^{i_{1}...j_{2n}}\Pi_{j_{1}...j_{2n}}
 \end{align}
 then, using (\ref{rosep22},\ref{s},\ref{ss}), we obtain
 \begin{align}\label{wegen}
 p_{j_{1},j_{2},...,j_{2n}}=\acute{p}_{j_{1},j_{2},...,j_{2n}}[1-x\lambda_{j_{1},j_{2},...,j_{2n}}^{w_{i_{1},i_{2},...,i_{2n}}}].
 \end{align}
 We now focus our attention on the subadditivity of correlations. By direct calculation one gets
$$
E+Q+C_{\sigma}-T_{\rho}=\sum  p_{j_{1},j_{2},...,j_{2n}}\log \frac{ p_{j_{1},j_{2},...,j_{2n}}}{\acute{p}_{j_{1},j_{2},...,j_{2n}}}+
\sum \acute{p}_{j_{1},j_{2},...,j_{2n}}\log \frac{\acute{p}_{j_{1},j_{2},...,j_{2n}}}{\acute{q}_{j_{1},j_{2},...,j_{2n}}}
$$
\begin{align}\label{nhae242}
+\sum \acute{q}_{j_{1},j_{2},...,j_{2n}}\log \acute{q}_{j_{1},j_{2},...,j_{2n}}-\sum p_{j_{1},j_{2},...,j_{2n}}\log p_{j_{1},j_{2},...,j_{2n}}=\sum (\acute{p}_{j_{1},j_{2},...,j_{2n}}-p_{j_{1},j_{2},...,j_{2n}})\log \acute{p}_{j_{1},j_{2},...,j_{2n}},
 \end{align}
 then, using (\ref{wegen}), we have
 \begin{align}\label{11111}
E+Q+C_{\sigma}-T_{\rho}=x\sum_{j_{1},j_{2},...,j_{2n}}\lambda_{j_{1},j_{2},...,j_{2n}}^{\phi_{i_{1},i_{2},...,i_{2n}}}\acute{p}_{j_{1},j_{2},...,j_{2n}}\log \acute{p}_{j_{1},j_{2},...,j_{2n}}
 \end{align}
 Assume, with no loss of generality, the $ w_{0,0,...,0}$, then we have $\acute{p}_{1,1,...,1}=\frac{2}{N^{2}}$, that is
 \begin{equation}\label{t5}
\sum_{j=1}^{2n}\acute{t}_{0,...,\underbrace{1}_{j},...,0}-(-1)^{n}\acute{t}_{1,1,...,1}+1=0.
\end{equation}
Thus, we have the following optimization problem
 \begin{equation}
\max \quad x\sum_{j_{1},j_{2},...,j_{2n}}\lambda_{j_{1},j_{2},...,j_{2n}}^{w_{0,0,...,0}}\acute{p}_{j_{1},j_{2},...,j_{2n}}\log \acute{p}_{j_{1},j_{2},...,j_{2n}}
 \end{equation}\label{tt5}
 \begin{equation}\label{lp2}
\rm{subject\; to}\left\{ \begin{array}{c}
\sum_{j=1}^{2n}(-1)^{i_{j}}\acute{t}_{0,...,\underbrace{1}_{j},...,0}+(-1)^{i_{2n+1}}\acute{t}_{1,1,...,1}\leq 1\\
\sum_{j=1}^{2n}\acute{t}_{0,...,\underbrace{1}_{j},...,0}-(-1)^{n}\acute{t}_{1,1,...,1}+1=0\\
\end{array}\right.
\end{equation}
The dual Lagrangian associated with this problem, is given by
$$
 L=x\sum_{j_{1},j_{2},...,j_{2n}}\lambda_{j_{1},j_{2},...,j_{2n}}^{\phi_{i_{1},i_{2},...,i_{2n}}}\acute{p}_{j_{1},j_{2},...,j_{2n}}\log \acute{p}_{j_{1},j_{2},...,j_{2n}}+\mu_{0,...,0}[\sum_{j=1}^{2n}\acute{t}_{0,...,\underbrace{1}_{j},...,0}-(-1)^{n}\acute{t}_{1,1,...,1}+1]
$$
\begin{equation}\label{11343}
+\sum_{i_{1},...,i_{2n+1}}\mu_{i_{1},...,i_{2n+1}}[\sum_{j=1}^{2n}(-1)^{i_{j}}\acute{t}_{0,...,\underbrace{1}_{j},...,0}+
(-1)^{i_{2n+1}}\acute{t}_{1,1,...,1}-1],
\end{equation}
and the complementary slackness condition(see the Appendix I) is given by
\begin{equation}\label{ppo}
\mu_{i_{1},...,i_{2n+1}}[\sum_{j=1}^{2n}(-1)^{i_{j}}\acute{t}_{0,...,\underbrace{1}_{j},...,0}+
(-1)^{i_{2n+1}}\acute{t}_{1,1,...,1}-1]=0
\end{equation}
The possible optimal solutions for above problem are the edge and vertices solutions.
First of all, we consider vertex solution that is one of the  vertices. In the case of odd n, we have
\begin{equation}\label{lpkl2}
\left\{\begin{array}{c}
  \acute{t}_{1,1,...,1}=- 1 \\
  \acute{t}_{0,...,\underbrace{1}_{j},...,0}=0, \quad  \forall j\in\{1,...,2n\}\
\end{array}\right.
\end{equation}
then $\acute{p}_{j_{1},j_{2},...,j_{2n}}=$ 0 if $j_{1}+j_{2}+...+j_{2n}$= odd and $\acute{p}_{j_{1},j_{2},...,j_{2n}}=\frac{2}{N^{2}}$ if $j_{1}+j_{2}+...+j_{2n}$= even.
Hence Eq.(\ref{11111}) gives
\begin{equation}\label{12443}
E+Q+C_{\sigma}-T_{\rho}=-\frac{(2n-1)x}{2^{3n-1}\sqrt{2(n+1)}}\log\frac{2}{N^{2}}\sum_{j_{1},j_{2},...,j_{2n}}[1+\sum_{k=1}^{2n}(-1)^{j_{k}}
 -(-1)^{j_{1}+j_{2}+...+j_{2n}}]=0
 \end{equation}
 Similarly, for other vertices( say $\acute{t}_{1,1,...,1}=- 1,\acute{t}_{0,...,\underbrace{1}_{j},...,0}=0, \quad  \forall j\in\{1,...,2n\}$(even n) and
, with no loss of generality, $\acute{t}_{1,0,...,0}=- 1 , \acute{t}_{1,1,...,1}=\acute{t}_{0,...,\underbrace{1}_{j},...,0}=0,
  \quad \forall j\in\{2,...,2n\}$) one can show that $E+Q+C_{\sigma}-T_{\rho}=0$.
Let us next turn our attention to the edge solutions. In this case we have
$$
 L=\sum_{i_{1},...,i_{2n}} p_{i_{1},...,i_{2n}}\log p_{i_{1},...,i_{2n}}-\sum_{i_{1},...,i_{2n}} p_{i_{1},...,i_{2n}}\log \acute{p}_{i_{1},...,i_{2n}}
$$
\begin{equation}\label{113443}
+\mu_{0,...,0}[\acute{t}_{10...0}+\acute{t}_{010...0}+...+\acute{t}_{00...1}-(-1)^{n}\acute{t}_{11...1}+1]
\end{equation}
By calculating the gradient of the dual Lagrangian with respect to making it zero one can show that
\begin{align}\label{ys}
\acute{t}_{10...0}=\acute{t}_{010...0}=...=\acute{t}_{00...1}.
 \end{align}

 Hence, using (\ref{rsep22},\ref{t5},\ref{ys}), one finds that
 $$
E+Q+C_{\sigma}-T_{\rho}=\frac{1}{2^{3n}\sqrt{2(2n+1)}}\sum_{i_{1}...i_{2n}}[1+\sum_{k=1}^{2n}(-1)^{i_{k}}-(-1)^{i_{1}+i_{1}+...
+i_{2n}}][1+(-1)^{i_{1}+i_{1}+...+i_{2n}}+
$$
\begin{align}\label{u123}
(\sum_{k=1}^{2n}(-1)^{i_{k}}+2n(-1)^{i_{1}+i_{1}+...+i_{n}})\acute{t}_{10...0}]\log\frac{1}{2^{2n}}[1+(-1)^{i_{1}+i_{1}+...+
i_{2n}}+(\sum_{k=1}^{2n}(-1)^{i_{k}}+2n(-1)^{i_{1}+i_{1}+...+i_{n}})\acute{t}_{10...0}]
 \end{align}
 The above function is convex and it is zero at $\acute{t}_{10...0}=0$, then one can immediately deduce that $E+Q+C_{\sigma}-T_{\rho}$
becomes negative for the acceptable value of $\acute{t}_{10...0}$.
 In general, the Eq.(\ref{11111}) is the difference of two convex functions, which both of them are non-positive and the intersection points of these two functions are in the vertices of the feasible region. On the other hand, for the  edge and vertices solutions $E+Q+C_{\sigma}-T_{\rho} \leq 0$. Then one can conclude that the correlations of generalized bloch sphere states are subadditive in the feasible region.



\section{Conclusion}
In the unified view of quantum and classical correlations the quantifications are done by the relative entropy and  optimization of relative
 entropy is known to be a difficult problem. In this work, we have presented a general algorithm via exact convex optimization to the problem of finding
CSS and CCS for a given entanglement state $\rho$. Using the obtained CCS for the $2^{n}$-dimensional Bell-diagonal states, we have derived analytical formula for the quantum discord of these states. As illustrating examples, we have analyzed the case of the separable generalized bloch sphere states and $2^{2}$-dimensional Bell-diagonal states and described how to apply the formula for this cases. We have also shown that the separable generalized bloch sphere states is subadditive. While our analysis is for a special states of bipartite quantum system, it serves to provide a unified explanation for a variety of states . In fact, this approach is completely general and could be applied for multipartite states in all dimensions. The main conclusion is that the presented algorithm provide
indispensable prerequisites for further investigation and can bring a robustness in constructing
CSS and CCS for a given multipartite states in all dimensions. Application of this algorithm to other quantum system and finding related
CSS and CCS is still an open problem which is under investigation.

\vspace{1cm} \setcounter{section}{0}
 \setcounter{equation}{0}
 \renewcommand{\theequation}{I-\arabic{equation}}
\newpage
{\Large{Appendix I:}}\\
\textbf{Convex optimization review:}
An optimization problem \cite{boyd}, has the standard form
\begin{equation}\label{conee}
\left\{ \begin{array}{c}
\mathrm{maximize} \ f_{0}(x) .\\
\mathrm{subject\; to}\quad
 \quad f_{i}(x)\leq b_{i}, i=1,...,m \quad h_{i}(x)=0, i=1,...,p\\
\end{array}\right.
\end{equation}
Where the vector $x= (x_{1}, ..., x_{n})$ is the optimization variable of the problem, the function $f_{0}:R^{n}\rightarrow R$
is the objective function, the functions $f_{i}:R^{n}\rightarrow R ,i=1,...,m$ are the (inequality) constraint functions, and the constants
$b_{1},...,b_{m}$ are the limits, or bounds, for the constraints. A convex optimization problem, is an optimization problem where the objective
and the constraint functions are convex functions which means they satisfy inequality $f_{i}(\alpha x+\beta y)\leq \alpha f_{i}(x)+ \beta f_{i}(y)$,
for all $(x,y,\alpha,\beta)\in R$ with $\alpha+\beta=1,\alpha\geq0,\beta\geq0$ and the equality constraint functions
$h_{i}(x)=0$  must be affine (A set $C\in R^{n}$ is affine if the line
through any two distinct points in C lies in C).
One can solve this convex optimization problem using Lagrangian duality. The basic idea in
the Lagrangian duality is to take the constraints in convex optimization problem into account
by augmenting the objective function with a weighted sum of the constraint functions. The
Lagrangian $L:R^{n}\times R^{m}\times R^{p}\rightarrow R$ associated with the problem is defined as
\begin{equation}\label{cofgn}
L(x,\lambda,\nu)=f_{0}(x)+\sum_{i=1}^{m}\lambda_{i}f_{i}(x)+\sum_{i=1}^{p}\nu_{i}h_{i}(x).
\end{equation}
The Lagrange dual function $g:R^{n}\times R^{m}\times R^{p}\rightarrow R$ is defined as the minimum value of
the Lagrangian over x: for $\lambda\in R^{m},\nu\in R^{p}$,
\begin{equation}\label{conhj1}
g(\lambda,\nu)=inf_{x\in D}L(x,\lambda,\nu).
\end{equation}

The dual function yields lower bounds on the optimal value $p^{\star}$ of the convex optimization
problem, i.e for any $\lambda\geq 0$ and any $\nu$ we have
\begin{equation}\label{conbn2}
g(\lambda,\nu)\leq p^{\star}.
\end{equation}
The optimal value of the Lagrange dual problem, which we denote $d^{\star}$, is, by definition, the
best lower bound on $d^{\star}$ that can be obtained from the Lagrange dual function. In particular,
we have the simple but important inequality
\begin{equation}\label{conghhg2}
d^{\star}\leq p^{\star}.
\end{equation}

This property is called weak duality. If the equality $d^{\star}= p^{\star}$ holds, i.e., the optimal duality
gap is zero, then we say that strong duality holds. If strong duality holds and a dual optimal
solution $(\lambda^{\star},\nu^{\star})$ exists, then any primal optimal point is also a minimizer of $L(x,\lambda^{\star},\nu^{\star} )$. This
fact sometimes allows us to compute a primal optimal solution from a dual optimal solution.
For the best lower bound that can be obtained from the Lagrange dual function one can
solve the following optimization problem
\begin{equation}\label{consdew}
\left\{ \begin{array}{c}
\mathrm{maximize} \ \ g(\lambda,\nu) .\\
\mathrm{subject\; to} \ \ \lambda\geq 0\\
\end{array}\right.
\end{equation}
This problem is called the Lagrange dual problem associated with the main problem. Conditions
for the optimality of a convex problem is called Karush-Kuhn-Tucker (KKT) conditions.
If $f_{i}$ are convex and $h_{i}$ are affine, and $\tilde{x},\tilde{\lambda},\tilde{\nu}$ are any points that satisfy the KKT conditions
$$
h_{i}(\tilde{x})=0, \ i=1,...,p,
$$
$$
f_{i}(\tilde{x})\leq0, \ i=1,...,m,
$$
$$
\tilde{\lambda}_{i}\geq 0 \ \ \tilde{\lambda}_{i}f_{i}(\tilde{x})=0, \ i=1,...,m,
$$
\begin{equation}
\bigtriangledown f_{0}(\tilde{x})+\sum_{i}^{m} \tilde{\lambda}_{i} \bigtriangledown f_{i}(\tilde{x})+\sum_{i}^{p}\tilde{\nu}_{i} \bigtriangledown h_{i}(\tilde{x})=0.
\end{equation}
then $\tilde{x}$ and $(\tilde{\lambda},\tilde{\nu})$ are primal and dual optimal, with zero duality gap. In other words, for
any convex optimization problem with differentiable objective and constraint functions, any
points that satisfy the KKT conditions are primal and dual optimal, and have zero duality
gap. Hence, $f_{0}(\tilde{x})=g(\tilde{\lambda},\tilde{\nu})$.\\
The condition $ \tilde{\lambda}_{i}f_{i}(\tilde{x})=0, \ i=1,...,m,$ is known as complementary slackness; it holds for
any primal optimal $x\tilde{}$ and any dual optimal $(\tilde{\lambda},\tilde{\nu})$ (when strong duality holds)
\vspace{1cm} \setcounter{section}{0}
 \setcounter{equation}{0}
 \renewcommand{\theequation}{I-\arabic{equation}}

{\Large{Appendix II:}}\\
Throughout the paper, we have used the formalism of Dirac $\gamma$ matrices. Therefore, in this appendix we define the algebra of Dirac $\gamma$ matrices and exhibit matrices which
realize the algebra in the Euclidean representation and explain our notations and conventions.\\
To do this, let $\gamma_{\mu}, \mu = 1, ..., d,$ be a set of d matrices satisfying the anticommuting relations:
\begin{align}\label{e}
\gamma_{\mu}\gamma_{\nu}+\gamma_{\nu}\gamma_{\mu}=2\delta_{\mu\nu}I,
\end{align}
in which I is the identity matrix.
These matrices are the generatores of a Clifford algebra similar to the algebra of operators
acting on Grassmann algebras. It follows from relations (\ref{e}) that the $\gamma$ matrices generate an
algebra which, as a vector space, has a dimension $2^{d}$. In the following, we will give an inductive
construction $(d \rightarrow� d+2)$ of hermitian matrices satisfying (\ref{e}). In the algebra one element
plays a special role, the product of all $\gamma$ matrices. The matrix $\gamma_{s}$:
\begin{align}
\gamma_{s}=i^{\frac{-d}{2}}\gamma_{1}\gamma_{2}...\gamma_{d},
\end{align}
anticommutes, because $d$ is even, with all other $\gamma$ matrices and $\gamma_{s}^{2}= I$.

In calculations involving $\gamma$ matrices, it is not always necessary to distinguish $\gamma_{s}$ from other
$\gamma$ matrices. Identifying thus $\gamma_{s}$ with $\gamma_{d+1}$, we have:
\begin{align}\label{ee}
\gamma_{i}\gamma_{j}+\gamma_{j}\gamma_{i}=2\delta_{ij}I, i,j=1,...,d,d+1.
\end{align}
The Greek letters $\mu\nu...$ are usually used to indicate that the value $d+1$ for the index has been
excluded.

\textbf{An explicit construction of $\gamma_{i}^{(d)}$}\\

It is sometimes useful to have an explicit realization of the algebra of $\gamma$ matrices.
For $d = 2$, the standard Pauli matrices realize the algebra:
$$
\gamma_{1}^{(d=2)}=\sigma_{1}=\left(
             \begin{array}{cc}
               0 & 1 \\
               1 & 0 \\
             \end{array}
           \right) , \gamma_{2}^{(d=2)}=\sigma_{2}=\left(
             \begin{array}{cc}
               0 & -i \\
               i & 0 \\
             \end{array}
           \right),
$$
\begin{align}
\gamma_{s}^{(d=2)}= \gamma_{3}^{(d=2)}=\sigma_{3}=\left(
             \begin{array}{cc}
               1 & 0 \\
               0 & -1 \\
             \end{array}
           \right)
\end{align}
The three matrices are hermitian, i.e., $\gamma_{i}=\gamma_{i}^{\dagger}$. The matrices $\gamma_{1}$ and $\gamma_{3}$ are symmetric and $\gamma_{2}$ is antisymmetric, i.e., $\gamma_{1}=\gamma_{1}^{t}$,  $\gamma_{3}=\gamma_{3}^{t}$ and $\gamma_{2}=-\gamma_{2}^{t}$.
To construct the � matrices for higher even dimensions, we then proceed by induction,
setting:
$$
\gamma_{i}^{(d+2)}=\sigma_{1}\otimes \gamma_{i}^{(d)}=\left(
                                                          \begin{array}{cc}
                                                            0 & \gamma_{i}^{(d)} \\
                                                            \gamma_{i}^{(d)} & 0 \\
                                                          \end{array}
                                                        \right), i=1,2,...,d+1,
$$
\begin{align}\label{eee}
\gamma_{d+2}=\sigma_{2}\otimes I^{(d)}=\left(
                                          \begin{array}{cc}
                                            0 & -iI_{d} \\
                                            iI_{d} & 0 \\
                                          \end{array}
                                        \right),
                                        \end{align}
where, $I_{d}$ is the unit matrix in $2^{\frac{d}{2}}$ dimensions.
As a consequence $\gamma_{s}^{(d+2)}$ has the form:
\begin{align}
\gamma_{s}^{(d+2)}=\gamma_{d+3}^{(d+2)}=\sigma_{3}\otimes I_{d}=\left(
                                                                    \begin{array}{cc}
                                                                      I_{d} & 0 \\
                                                                      0 & -I_{d} \\
                                                                    \end{array}
                                                                  \right)
\end{align}
A straightforward calculation shows that if the matrices $\gamma_{i}^{(d)}$ satisfy relations (\ref{ee}), the $\gamma_{i}^{(d+2)}$ matrices satisfy the same relations. By induction we see that the $\gamma$ matrices are all hermitian. from (\ref{eee}), it is seen that, if $\gamma_{i}^{(d)}$ is symmetric or antisymmetric, $\gamma_{i}^{(d+2)}$
 has the same property.
 The matrix $\gamma_{d+2}^{(d+2)}$ is antisymmetric and the matrix  $\gamma_{d+3}^{d+2}$
 is symmetric. It follows immediately that, in this representation, all $\gamma$ matrices with odd index are symmetric and all matrices with even index are antisymmetric, i.e.,
\begin{align}
\gamma_{i}^{t}=(-1)^{i+1}\gamma_{i}.
\end{align}

\end{document}